\documentclass[aps,twocolumn,superscriptaddress,floatfix]{revtex4-1}
\usepackage{graphicx,amsfonts,amssymb,amsmath,mathrsfs,hyperref,bm,dsfont}

\newif\ifhyper
\hypertrue
\ifhyper
\hypersetup{
   citecolor = {red},
   colorlinks = {true}, 
   linkcolor = {blue},
   urlcolor = {blue} 
}
\fi

\def\be{\begin{equation}}
\def\ee{\end{equation}}
\def\bea{\begin{eqnarray}}
\def\eea{\end{eqnarray}}

\newcommand{\Zd}{\mathbb{Z}_2}

\newcommand{\ket}[1]{|#1\rangle}
\newcommand{\bra}[1]{\langle #1|}

\newcommand{\mc}[1]{\mathcal{#1}}

\newcommand{\resloopa}{ {\parbox[c]{0.1\linewidth}{\includegraphics[width=\linewidth]{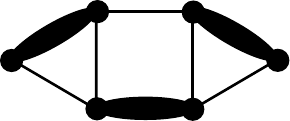} } }}
\newcommand{\resloopb}{ {\parbox[c]{0.1\linewidth}{\includegraphics[width=\linewidth]{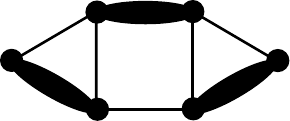} } }}

\newcommand{\resflipa}{ {\parbox[c]{0.1\linewidth}{\includegraphics[width=\linewidth]{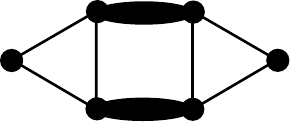} } }}
\newcommand{\resflipb}{ {\parbox[c]{0.1\linewidth}{\includegraphics[width=\linewidth]{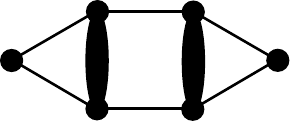} } }}

\newcommand{\nores}{ {\parbox[c]{0.08\linewidth}{\includegraphics[width=\linewidth]{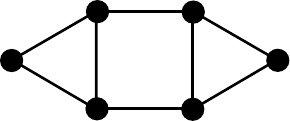} } }}

\begin{document}

\title{Topological $\mathbb{Z}_2$ RVB quantum spin liquid on the ruby lattice }

\author{Saeed S. Jahromi}
\email{saeed.jahromi@dipc.org}
\affiliation{Donostia International Physics Center, Paseo Manuel de Lardizabal 4, E-20018 San Sebasti\'an, Spain}

\author{Rom\'an Or\'us}
\affiliation{Donostia International Physics Center, Paseo Manuel de Lardizabal 4, E-20018 San Sebasti\'an, Spain}
\affiliation{Ikerbasque Foundation for Science, Maria Diaz de Haro 3, E-48013 Bilbao, Spain}
\affiliation{Multiverse Computing, Pio Baroja 37, 20008 San Sebasti\'an, Spain}

\begin{abstract}

We construct a short-range resonating valence-bond state (RVB) on the ruby lattice, using projected entangled-pair states (PEPS) with bond dimension $D=3$. By introducing non-local moves to the dimer patterns on the torus, we distinguish four distinct sectors in the space of dimer coverings, which is a signature of the topological nature of the RVB wave function. Furthermore, by calculating the reduced density matrix of a bipartition of the RVB state on an infinite cylinder and exploring its entanglement entropy, we confirm the topological nature of the RVB wave function by obtaining non-zero topological contribution, $\gamma=-\rm{ln}\ 2$, consistent with that of a $\Zd$ topological quantum spin liquid. We also calculate the ground-state energy of the spin-$\frac{1}{2}$ antiferromagnetic Heisenberg model on the ruby lattice and compare it with the RVB energy. Finally, we construct a quantum-dimer model for the ruby lattice and discuss it as a possible parent Hamiltonian for the RVB wave function. 

\end{abstract}

\maketitle

\section{Introduction.}
 
Quantum spin liquids (QSL) \cite{Savary2017} are intriguing phases of matter which emerge in different systems ranging from quantum antiferromagnets \cite{Balents2010,Liao2017,Poilblanc2019,Jahromi2018a} to superconducting phases \cite{Anderson1987,Poilblanc2014} and topologically ordered systems \cite{Wen1990,Wen1995,Levin2005,Kitaev2003,Kitaev2006,Jahromi2013a,Jahromi2013,Jahromi2017,Jahromi2016,Mohseninia2015,Capponi2014}. Due to strong zero-point quantum fluctuations and lack of conventional magnetic long-range order, they are very difficult to detect and their experimental realization is still a challenge \cite{Savary2017}. However, their exotic characteristics such as highly entangled ground-states, fractional excitations \cite{Kitaev2003,Kitaev2006,Levin2005}, and topological properties \cite{Levin2006,Kitaev2006a,Jahromi2017} strongly motivate the study of such phases of quantum matter. 

Resonating valence-bond states are the very first instances of QSL. They were introduced by Anderson \cite{Anderson1987} as the possible ground state of the spin-$1/2$ Heisenberg antiferromagnet on the triangular lattice \cite{Fazekas1974}, and later as the parent Mott insulating state in the context of high-temperature superconductivity \cite{Bednorz1986}. The RVB wave function is formed by first covering all the lattice with spin-zero singlets between nearest-neighboring spins, and then resonating (i.e. equal-weight adding) over all possible such coverings of the lattice \cite{Lacroix2013}. This superposition is highly entangled and breaks no symmetries \cite{Poilblanc2012,Poilblanc2013}. The RVB state is, therefore, a candidate to describe the physics of quantum spin liquids \cite{Poilblanc2013,Wang2013a,Gauthe2019,Chen2018a,Chen2018}. 

The RVB states are typically expected as the underlying ground states in quantum antiferromagnetic $\rm{SU(2)}$ symmetric models \cite{Lacroix2013} such as frustrated spin-$\frac{1}{2}$ Heisenberg model \cite{Poilblanc2019,Iqbal2011}. Different numerical simulations already suggested the RVB states as the ground state of the spin-$\frac{1}{2}$ antiferromagnetic Heisenberg model on the kagome lattice \cite{Iqbal2014,Iqbal2011,Yan2011,Depenbrock2012,Jiang2012}. Besides, the RVB wave functions have also been exploited as the underlying theory for describing the resonance mechanism in the Rokhsar-Kivelson quantum dimer models \cite{Rokhsar1988} on the square \cite{Moessner2001,Syljuasen2006} and triangular lattices \cite{Moessner2002,Ralko2005}.   

Recently, the tensor network (TN) representation of RVB states has been put forward \cite{Schuch2012} and exact TN representation for RVB state on the square lattice \cite{Chen2018,Poilblanc2019,Wang2013a} and topological $\Zd$ RVB states on the kagome lattice for both nearest-neighbor \cite{Schuch2012,Poilblanc2012} and next-nearest-neighbor dimer coverings \cite{Poilblanc2013} have been introduced. Moreover, the TN construction of the RVB wave function has been applied to chiral spin liquids \cite{Chen2018a,Poilblanc2016,Poilblanc2015} as well as semionic \cite{Iqbal2014} and fermionic \cite{Poilblanc2014} RVB states. Besides, there have been several proposals for RVB ansatz in the context of $\rm{SU(N)}$ models \cite{Gauthe2019,Poilblanc2016}. 

\begin{figure}
\centerline{\includegraphics[width=7cm]{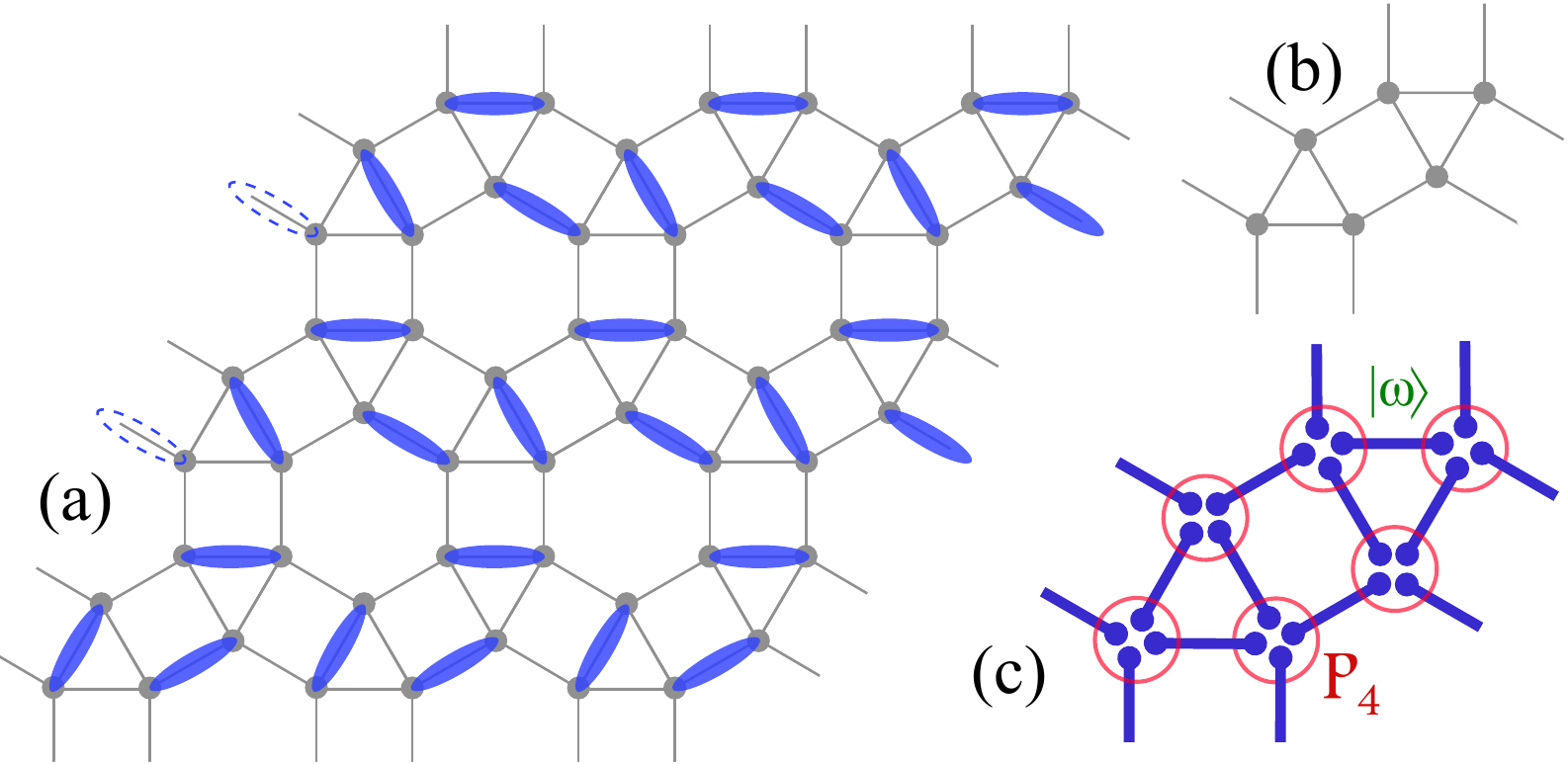}}
\caption{(Color online) (a) An example of a dimer covering on the ruby lattice. (b) The unit-cell of the ruby lattice composed of two triangles with six sites. (c) Covering of the ruby unit-cell with entangled state $\ket{\omega}$ and the projector $\mc P_4$ which maps the virtual degrees of freedom to the local physical space.}
\label{Fig:ruby}
\end{figure}

While the RVB ansatz has played a key role in understanding the physics of QSL, there are still plenty of systems and lattice geometries that are not yet fully explored. The ruby lattice, also known as bounce lattice or $3.4.6.4$ geometry (see Fig.~\ref{Fig:ruby}-(a)) \cite{Jahromi2016,Jahromi2018,Kargarian2010}, is one of the structures from the family of Archemedian lattices \cite{Farnell2014} with potential for hosting QSLs. It has already been shown that the anisotropic Kitaev model on the ruby lattice hosts two gapless and a gaped spin liquid phase \cite{Jahromi2016,Jahromi2018} in which the latter has been proven to be $\Zd\times\Zd$ topologically ordered \cite{Kargarian2010,Bombin2009}. It has further been shown that the low-energy effective theory of the gaped QSL phase is described by a topological color code (TCC) \cite{Bombin2006} with a topologically degenerate ground state and anyonic excitations, thus being a suitable platform for fault-tolerant quantum computation. The TCC is, in fact, two coupled copy of the toric code (TC) \cite{Kubica2015,Bombin2012,Bombin2014} and belongs to the same family of abelian theories as that of the TC \cite{Levin2005,Kitaev2003}. Furthermore, the ruby lattice is also of experimental relevance:  the bismuth ions in layered materials such as $\rm{Bi_{14}Rh_3I_9}$ form ruby structure with interesting topological properties \cite{Rasche2013,Pauly2015,Pauly2016}. 

In this paper, we introduce a $\Zd$ topologically ordered RVB wave function on the ruby lattice as a $D=3$ projected entangled-pair state to better understand the topological nature of QSL states on this lattice. In order to reveal the topological nature of the state, we first distinguish four distinct sectors in the space of dimer coverings on the torus and then calculate the reduced density matrix of the RVB state for a semi-infinite cylinder, and extract from it the non-zero topological contribution to the entanglement entropy, in turn characterizing topological order. We further elaborate on the possible parent Hamiltonians for the RVB state on the ruby lattice and discuss its relation to the $\Zd$ topological order of the toric code, as well as compare the energy of the RVB to that of the ground state of the Heisenberg model for this lattice. 

The paper is organized as follows. In Sec.~\ref{sec:PEPS-RVB} we introduce two different representations for the RVB state on the ruby lattice in TN language. Next, in Sec.~\ref{sec:To-Sectors} we show how to construct the four topologically degenerate RVB states on the torus. In Sec.~\ref{sec:To-Entropy} We calculate the RDM of the RVB wave function on the semi-infinite cylinder and extract the topological entropy of the state. We further elaborate on the possible parent Hamiltonians for the RVB state in Sec.~\ref{sec:Par-Ham}. Finally Sec.~\ref{sec:conclude} is devoted to the discussion and conclusions.

\section{Construction of the RVB Wave Function} 
\label{sec:PEPS-RVB}  

Before we elaborate on the construction of the RVB wave function on the ruby lattice, let us first briefly review the basic concepts of tensor network states and in particular the projected entangled-pair states.

\subsection{Projected entangled-pair state}
Tensor network states are efficient representations of many-body wave functions which codify the states according to their entanglement structure \cite{Orus2014,Orus2019,Ran2017,Biamonte2017,Verstraete2008}. The matrix product state (MPS) \cite{Fannes1992,Ostlund1995} is an example of a 1D TN state and has been proved to provide a very efficient representation for the ground states of 1D gapped local Hamiltonians. The MPS has also been the heart of 1D numerical TN methods such as density matrix renormalization group (DMRG) \cite{White1992,White1993} and time-evolving block decimation (TEBD) \cite{Vidal2004}. The extension of MPS to 2D and higher dimensions has also been put forward in the form of projected entangled-pair state (PEPS) \cite{Vidal2007,Corboz2010,Phien2015}. During past years, PEPS has been successfully used for representing the ground state of numerous quantum many-body systems and has played a key role in a better understanding of strongly correlated systems \cite{Orus2009,Corboz2013,Matsuda2013,Corboz2014,Corboz2014a,Jahromi2018a,Jahromi2018,Jahromi2019,Schmoll2019}. 

PEPS are states which can be described by first placing ``virtual" entangled states between the sites of the system, and subsequently applying linear maps at each site to obtain the physical system. The dimension of the virtual degree of freedom is the bond dimension $D$, which is a measure of the amount of entanglement that can be handled by the system. The many-body wave function is then constructed by contracting (tensor-tracing) all the virtual degrees of freedom on the given lattice, i.e
\be
\ket{\Psi}=\sum_{\{s_{\vec{r}_i}\}_{i=1}^N}^d \mathcal{F} \left( A_{s_{\vec{r}_1}}^{[\vec{r}_1]}, \ldots, A_{s_{\vec{r}_N}}^{[\vec{r}_N]} \right) \ket{s_{\vec{r}_1}, \ldots, s_{\vec{r}_N}},
\ee
where $\ket{s_{\vec{r}_i}}$ is the local basis of the site $i$ at position 
$\vec{r}_i$ according to the geometry of the lattice and 
$A_{s_{\vec{r}_i}}^{[\vec{r}_i]}$ are the local tensors or linear maps from the physical to virtual spaces \cite{Verstraete2008,Orus2014,Orus2019}. In what follows, we introduce two different (but equivalent) constructions for the RVB state on the ruby lattice within the PEPS formalism. 

\begin{figure}  
\centerline{\includegraphics[width=7cm]{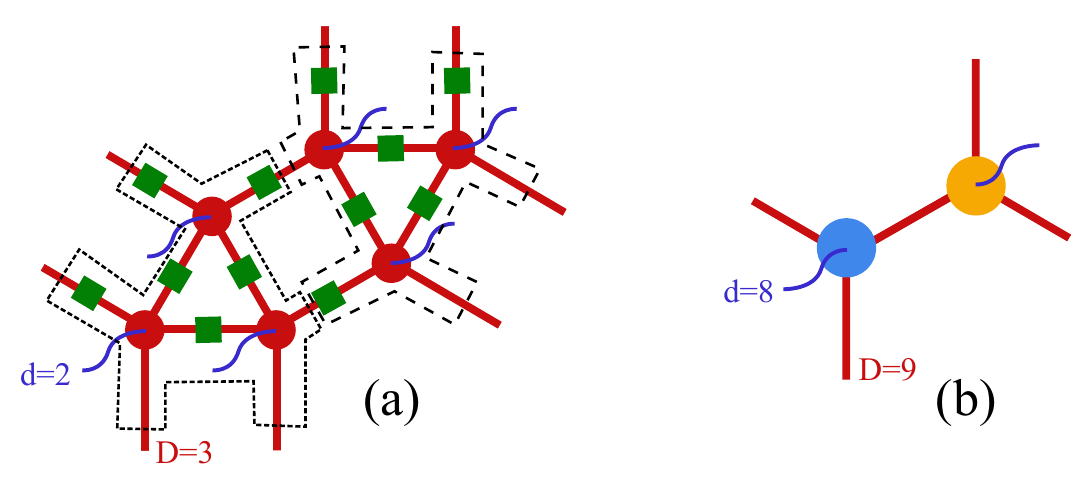}}
\caption{(Color online) (a) Coarse-graining every three tensors at the vertices of the triangles of the ruby uni-cell to a block tensor with physical dimension $d=2^3$ and virtual dimension $D=9$. (b) The resulting coarse-grained tensors form a honeycomb TN.}
\label{Fig:ruby-block}
\end{figure}

 \subsection{PEPS representation of the RVB state}

In this section, we construct the nearest-neighbour (NN) RVB state on the ruby lattice. The NN RVB is constructed by an equal-weight superposition of all dimer coverings with hard-core constraint \cite{Lacroix2013}, i.e, every site of the lattice can contribute to one and only one NN dimer (singlet) on the ruby lattice. This implies that every link of the lattice can be either empty or occupied by a singlet according to the underlying dimer covering. In order to obtain the PEPS representation of the RVB state, we first place the following maximally-entangled state of two qutrits on each link of the ruby lattice: 
 \begin{equation}
        \label{eq:rvb-bond}
\ket{\omega} = \ket{01}-\ket{10}+\ket{22}. 
\end{equation}
This is shown on Fig.\ref{Fig:ruby}-(c) for the ruby unit-cell. The above state of two qutrits is nothing but the conventional $\rm{SU(2)}$ singlet plus an empty state $\ket{22}$ for representing links with no singlets. Next, we act on every site of the lattice with the projector
\begin{equation}
\label{eq:rvb-peps-proj}
\begin{aligned}
\mc P_4 =& \ket{0}\big(\bra{0222}+\bra{2022}+\bra{2202}+\bra{2220}\big)\\
     &+\ket{1}\big(\bra{1222}+\bra{2122}+\bra{2212}+\bra{2221}\big)\end{aligned}
\end{equation}
 to map every site $j$ of the lattice to a two-level Hilbert space (qubit). The RVB wave function on the ruby lattice is therefore given by  
\be
\label{eq:RVB1}
\ket{\Psi}=\prod_j \mc P_4^j \prod_{\langle i,j \rangle} \ket{\omega}.
\ee 

The intuition behind the mapping in Eq.~\eqref{eq:rvb-peps-proj} is to ensure the hard-core constraint such that every site participates in the formation of a singlet $\ket{01}-\ket{10}$ along each link only once (being in the states $\{\ket{01}, \ket{10}\}$), while the rest of the links are unoccupied (they are in the states $\ket{22}$). This condition is further satisfied by the specific form of the entangled state $\ket{\omega}$, which ensures that if one of the sites of a link is in the state $\ket{0}$, then the second one must be in the state $\ket{1}$ and vice versa (otherwise both of them are in the state $\ket{2}$ and then the link is empty). On a 2D lattice, the action of the projector $\mc P_4$ on state $\ket{\omega}$ gives thus rise to an equal-weight superposition of all possible dimer coverings, which is the requirement of an RVB state. 

 \begin{figure}  
\centerline{\includegraphics[width=7cm]{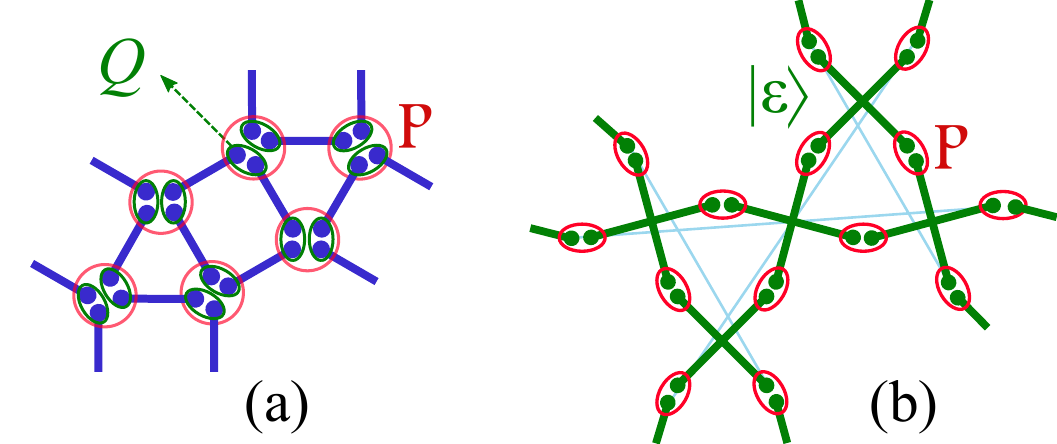}}
\caption{(Color online) (a) Covering of the ruby unit cell with the entangled state $\ket{\omega}$ and the projectors $\mc P$ acting on the four $\ket{\omega}$ states around each square of the ruby lattice and constructing the simplex tensor $\ket{\varepsilon}$. (b) The green simplex tensors together form a kagome lattice. The light blue line shows the underlying kagome grid which is obtained by squeezing the green simplices.}
\label{Fig:simplex}
\end{figure}

In order to study the properties of the RVB wave function \eqref{eq:RVB1} in the thermodynamic limit, one has to contract the state on an infinite 2D ruby lattice. Since the exact contraction of an infinite 2D PEPS is a computationally expensive task \cite{Orus2014,Ran2017}, one has to use approximation methods such as tensor renormalization group (TRG) \cite{Levin2007, Gu2008}, boundary MPS methods (BMPS) \cite{Vidal2007} or corner transfer-matrix renormalization group (CTMRG) \cite{Orus2009,Corboz2014a,Corboz2010a,Nishino1996}. The standard approach for performing such approximation methods is to coarse-grain several lattice sites into a network of block sites with a square geometry in order to handle the renormalization process very efficiently \cite{Jahromi2018a,Jahromi2018}. Coarse-graining the ruby lattice into block-sites involves grouping every three sites of a ruby triangle into a coarse-grained local tensor with a physical dimension of $d=2^3=8$ and three virtual bonds with $D=9$, which together form a honeycomb TN (see Fig.~\ref{Fig:ruby-block}). Although contracting a tensor network with $D=9$ is still tractable, subject to the availability of sufficient memory and CPU time, studying the thermodynamic properties of the ruby RVB state within such a setting is a computationally demanding task.

In the next subsection, we introduce a computationally more efficient representation for the RVB state on the ruby lattice which involves the contraction of a TN with bond dimension $D=3$.          

\subsection{Simplex RVB representation}

\begin{figure}  
\centerline{\includegraphics[width=7cm]{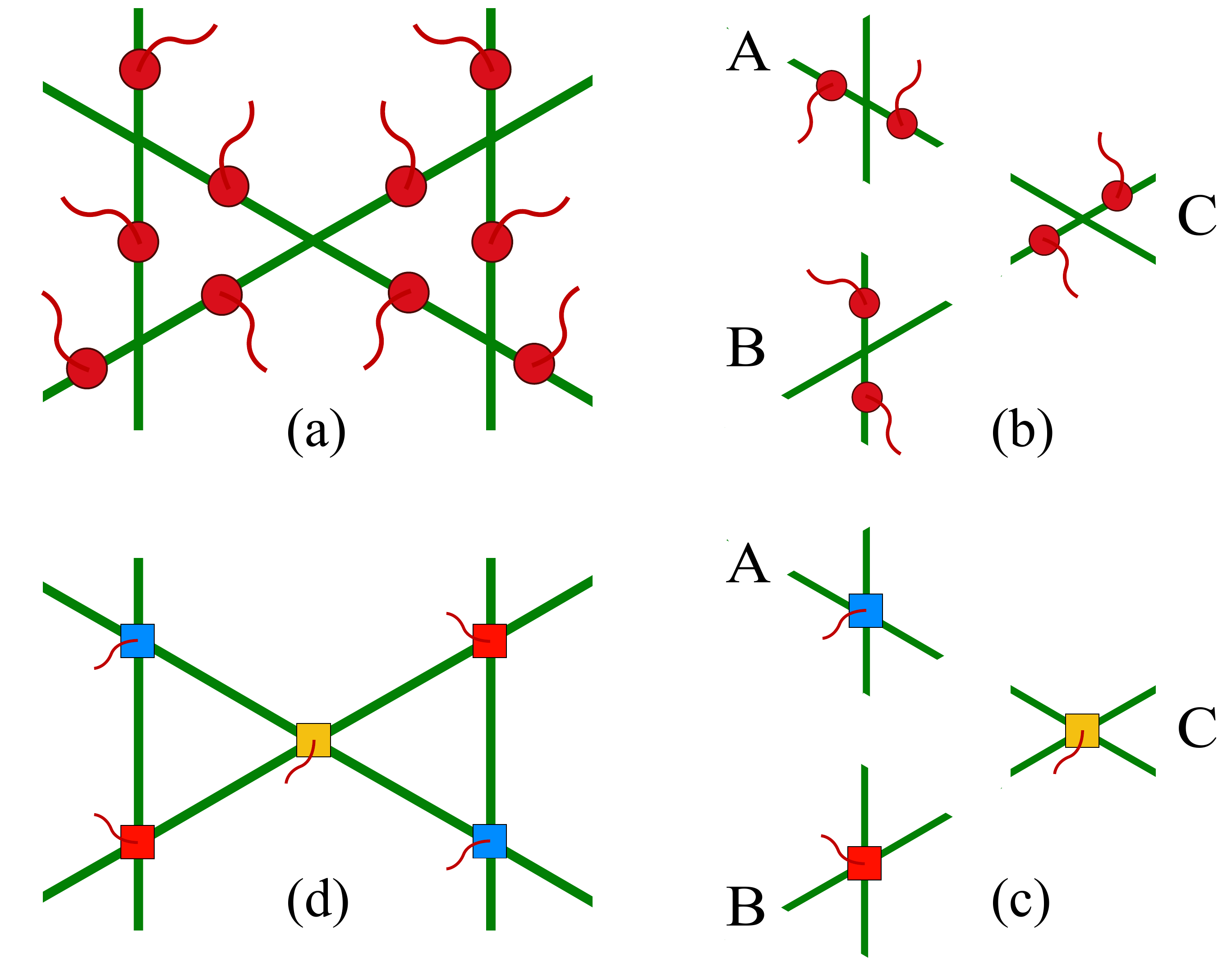}}
\caption{(Color online) (a) Kagome lattice formed by rank-4 simplex tensors and rank-3 projectors $\mc P$. (b) Absorbing two projectors  $\mc P$ into every simplex tensor one obtains three new tensors $A$, $B$ and $C$ with $d=4$ and $D=3$ shown in (c) which span the kagome lattice (d).}
\label{Fig:block}
\end{figure}

In the second construction of the RVB wave function, we again start by placing the entangled state $\ket{\omega}$, Eq.~\eqref{eq:rvb-bond}, on every link of the ruby lattice. Next, we redefine the projector $\mc P_4$ in the following way:
\begin{equation}
\label{eq:P-P2Q}
\mc P_4 = \mc P\;(\mc Q\otimes \mc Q)\ .
\end{equation} 
The new projector $\mc P$ is defined in a similar fashion as 
as $\mc P_4$. However, as shown in Fig.~\ref{Fig:simplex}-(a) now it acts only on two virtual systems,
\begin{equation}
	\mc P = \ket{0}(\bra{02}+\bra{20})+\ket{1}(\bra{12}+\bra{21}),
\end{equation}
and $\mc Q$ is defined as
\begin{equation}
\mc Q = \ket{0}\big(\bra{02}+\bra{20}\big)+
    \ket{1}\big(\bra{12}+\bra{21}\big) + 
    \ket{2}\bra{22}.\ 
\end{equation}
Applying $\mc Q^{\otimes 4}$ to $\ket{\omega}^{\otimes 4}$ across each square of the ruby lattice yields a $4$-qutrit state $\ket{\varepsilon_{ijkl}}$, which is a symmetric simplex tensor representing the four-partite entanglement structure of the RVB state inside a square part of the ruby lattice. Its non-zero elements are given in Table~\ref{tab1}. The RVB wave function is then constructed by acting with the projector $\mc P$ on the ruby lattice as depicted in Fig.~\ref{Fig:simplex}-(b). 

The huge advantage of the simplex RVB construction over the previous one is that the simplex tensors $\ket{\varepsilon_{ijkl}}$ form a kagome lattice, as shown in Fig.~\ref{Fig:simplex}-(b) and \ref{Fig:block}. The rank-4 green tensors in Fig.~\ref{Fig:block}-(a) are the simplex tensors which coincide on the vertices of the kagome lattice and the rank-3 red tensors are the projectors $\mc P$ which act on the links of the kagome lattice. Since the kagome lattice is a tripartite lattice, one can absorb two $\mc P$ projectors into every simplex tensors as shown in Fig.~\ref{Fig:block}-(b,c) to obtain three new tensors $A$, $B$ and $C$ with $d=4$ and $D=3$. These tensors span the kagome lattice with a three-site translationally invariant unit-cell as shown in Fig.~\ref{Fig:block}-(d). This yields a $D=3$ kagome tensor network which can be contracted very efficiently with, e.g., TRG or BMPS (see also Appendix~\ref{app:BMPS}).   

\begin{table}[!t]
 \begin{center}
 \caption{Non-zero elements of the simplex tensor $\ket{\varepsilon}$ used in the construction of the RVB state on the ruby lattice.}
 \label{tab1}
\begin{ruledtabular}
	\begin{tabular}{l*{3}{c}r}
 $\varepsilon(1,2,1,2)=2$,\quad\quad  & $\varepsilon(2,1,2,1)=2$,\quad\quad    & $\varepsilon(1,2,3,3)=1$ \\ 
 $\varepsilon(2,3,3,1)=1$,\quad\quad  & $\varepsilon(3,1,2,3)=1$,\quad\quad  & $\varepsilon(3,3,1,2)=1$ \\  
 $\varepsilon(3,3,3,3)=1$,\quad\quad  & $\varepsilon(1,1,2,2)=-1$,\quad\quad & $\varepsilon(1,2,2,1)=-1$ \\  
 $\varepsilon(1,3,3,2)=-1$,\quad\quad & $\varepsilon(2,1,1,2)=-1$,\quad\quad & $\varepsilon(2,1,3,3)=-1$ \\  
 $\varepsilon(2,2,1,1)=-1$,\quad\quad & $\varepsilon(3,2,1,3)=-1$,\quad\quad & $\varepsilon(3,3,2,1)=-1$ 
	\end{tabular}
\end{ruledtabular}
 \end{center}
\end{table}

\section{Topological Sectors} 
\label{sec:To-Sectors} 

\begin{figure}   
 \centerline{\includegraphics[width=6cm]{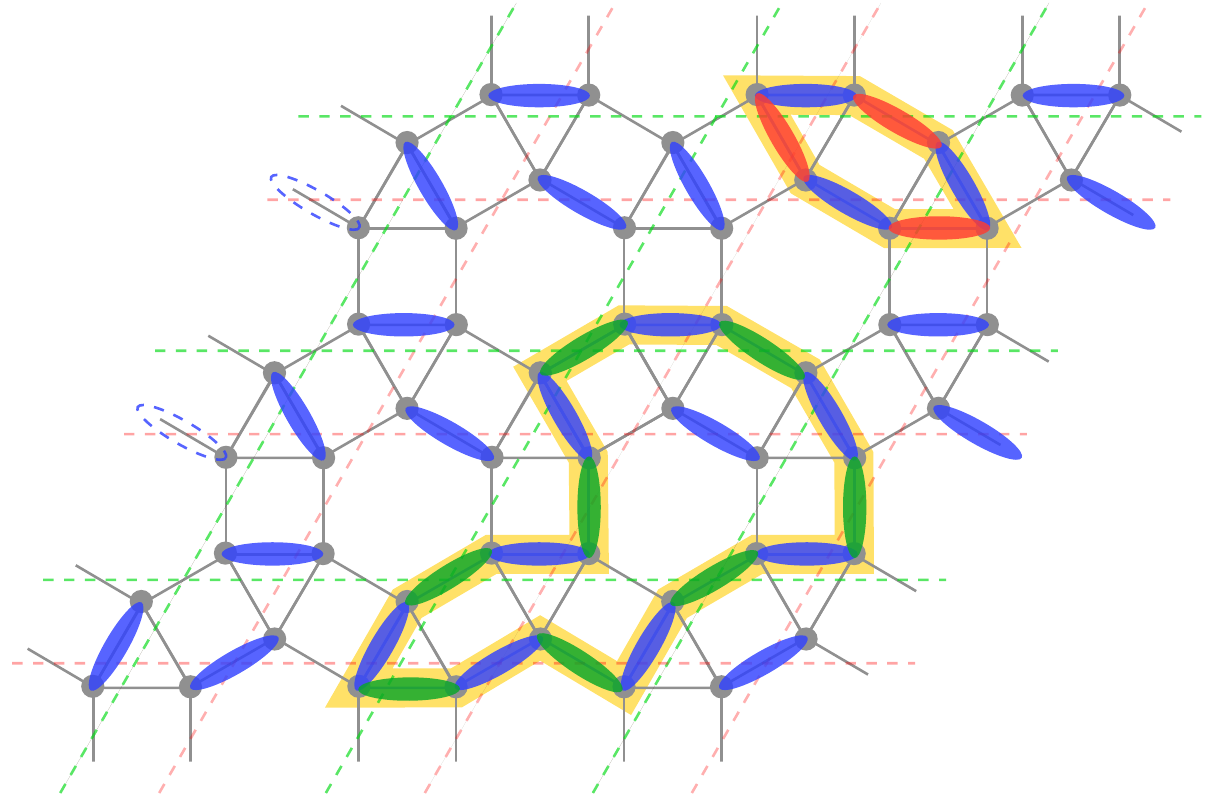}}
\caption{(Color online) Examples of two local resonances on a dimer covering. The dimers can resonate between the two configurations distinguished by blue and red (green) along the highlighted yellow loops. The dashed lines are non-contractible reference lines to specify the winding numbers in each topological sector.}
\label{Fig:resonance}
\end{figure}

In contrast to standard spin models, where different configurations in the Hilbert space can be constructed by simply flipping single spins, the hard-core constraint in the dimer model prohibits such movements to assure that every spin is participating in the formation of one and only one singlet on a specific link of the underlying lattice. As a consequence of this constraint, it is therefore not possible to manipulate a dimer without moving other dimers. The simplest moves on the ruby lattice would, therefore, involve shifting the dimers along flippable loops such as a ruby diamond-plaquette as depicted in Fig.~\ref{Fig:resonance}.  Generalization to larger loops is obviously possible.

Within such a restricted Hilbert space, there are quantities that remain invariant under the local movements of dimers, such as the parity of the number dimers crossed by a non-contractible reference line. Examples of such reference lines are illustrated in Fig.~\ref{Fig:resonance}. One can clearly check that every reference line always crosses an even number of dimers and any local rearrangements of dimers such as those along the highlighted loops in the figure would not change the parity.  

\begin{figure*}
\centerline{\includegraphics[width=18cm]{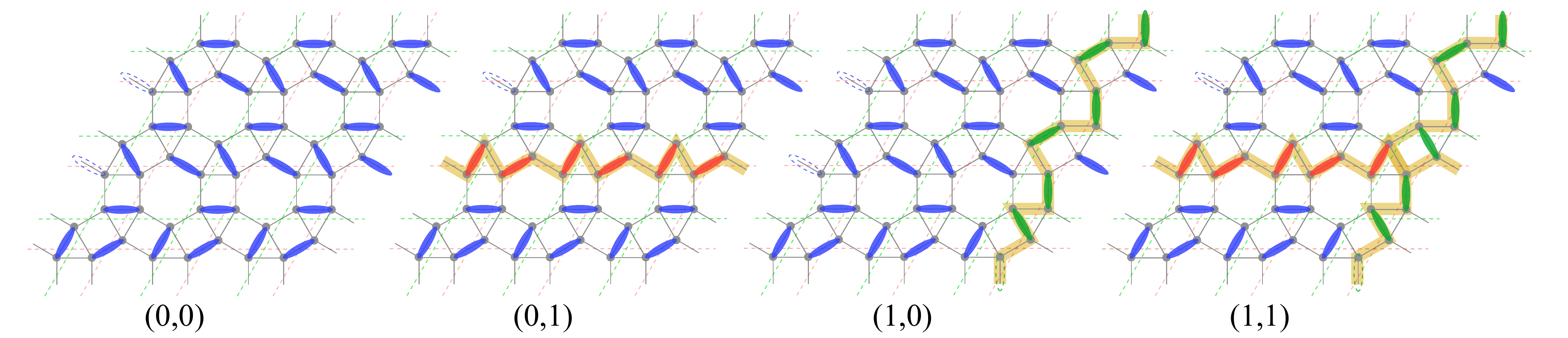}}
\caption{(Color online) Examples of the dimer covering on the four topological sectors of the ruby RVB wave function on the torus with genus $g=1$. The topological sectors are characterized by winding numbers  ($W_h$,$W_v$) which are the parity of the number of dimers crossed by non-contractible horizontal and vertical reference lines (dashed green and red lines). One can interchange between the sectors by moving the dimers along the non-local loops around the toroidal directions (highlighted by yellow path).}
\label{Fig:ToSec}
\end{figure*}

On the contrary, if the lattice is warped around a torus or a manifold of a higher genus, it is possible to find paths along the toroidal directions such that if the dimers are moved along them, then the parity changes. Fig.~\ref{Fig:ToSec} shows examples of such non-contractible loops around the torus for the ruby lattice. Defining the winding numbers ($W_h$,$W_v$) for characterizing the parity of the dimer crossings for the two homology classes of the torus in the horizontal and vertical directions, one can label the four possible sectors for the RVB wave function on a torus with genus $g=1$. It is clear that any local move in the dimer pattern would preserve the parity of the state. However, interchanging between the sectors would be possible by introducing non-local moves along the toroidal directions. Examples of such non-local moves are illustrated in Fig.~\ref{Fig:ToSec}. The left-most dimer configuration in the figure has even parity in both horizontal and vertical directions given by ($0,0$) winding numbers. By rearranging the dimers along the global loops shown in the figure, one can change the winding numbers along each toroidal direction until reaching the odd parity sector, ($1,1$), on the right side of the figure.

These sectors are the consequence of the topology of the surface and therefore are called {\it topological} sectors \cite{Kitaev2003,Schuch2012,Poilblanc2012}. They are the subspace spanned by the set of all dimer coverings with a given winding number. For more general surfaces of higher genus, the sectors are characterized by specifying $2g$ winding numbers. In the next section, we provide another signature to characterize the topological order of the RVB wave function on the ruby lattice.         

\section{Topological Entropy} 
\label{sec:To-Entropy} 

In order to find a more concrete evidence on the topological nature of the RVB state on the ruby lattice, in this section, we study the entanglement properties of the RVB wave function. In particular, we evaluate the entanglement entropy (EE) of the RVB state by calculating the reduced density matrix (RDM) on semi-infinite cylinders with finite width $N_v$ and infinite height $N_h\longrightarrow\infty$. 

We specifically calculate the Von Neumann EE given by $S=-\sum_i \lambda_i \ {\rm ln} \ \lambda_i$,where $\lambda_i$ are the spectrum of the RDM \cite{Nielsen:2011:QCQ:1972505}. It has already been known that the EE of 2D gaped phases with topological order scales as $S=\alpha L+\gamma+O(1/L)$ where the first part is a linear term arises from the area-law scaling of the EE with the boundary of the region, $L$, and the second term, $\gamma$, gives access to the total quantum dimension $\mc D$ which is a strong fingerprint of the topological order \cite{Levin2006,Kitaev2006a}.   

\begin{figure}  
  \centerline{\includegraphics[width=\columnwidth]{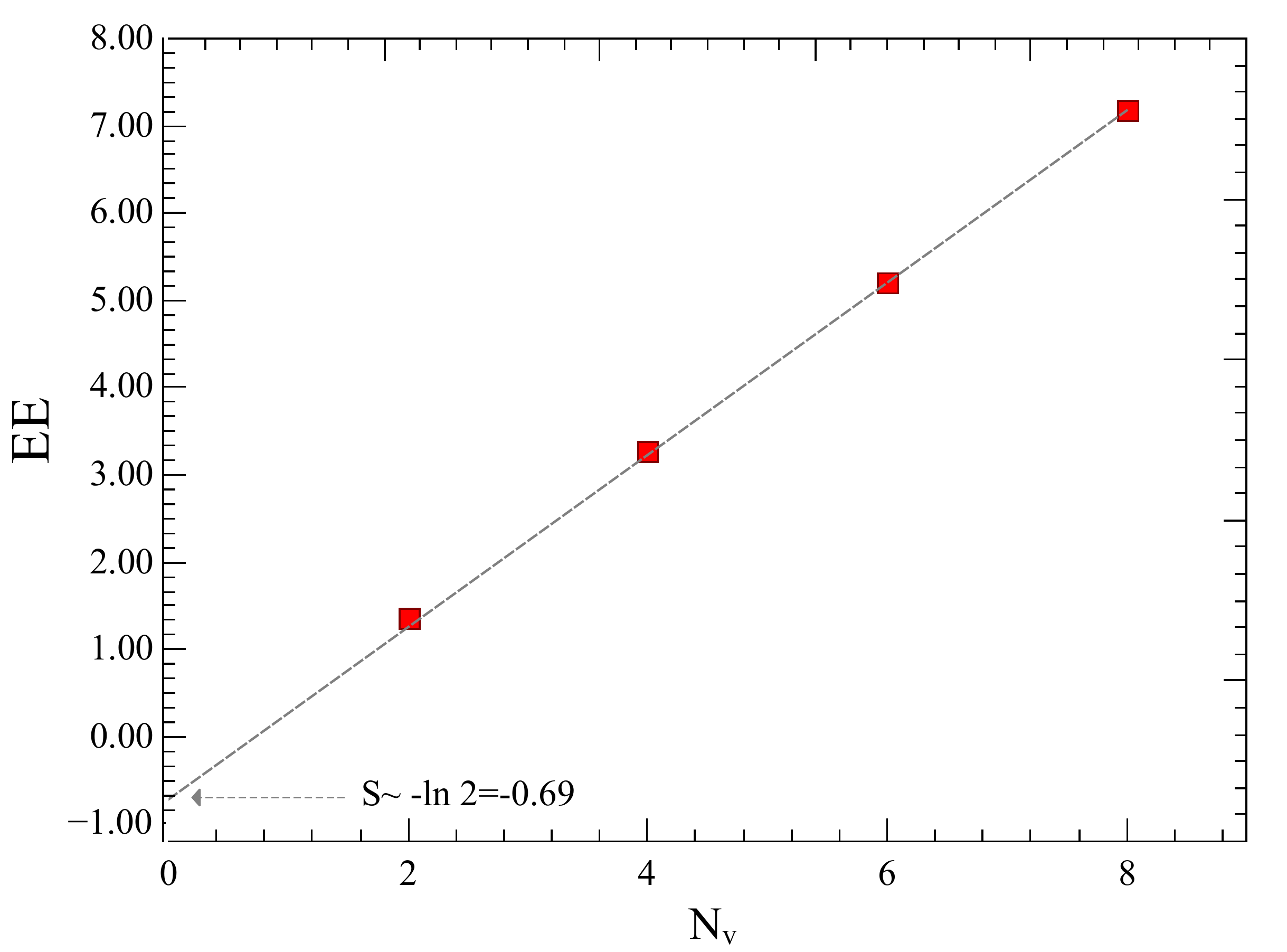}}
\caption{(Color online) Entanglement Entropy of the RVB wave function on the ruby lattice versus the cylinder width $N_v$ in the even sector, ($0,0$). The dashed line is the linear fit using the two largest cylinder perimeters, i.e, $N_v=6,8$.}
\label{Fig:Svn}
\end{figure}

In order to obtain the EE of the RVB state on the cylinder, we calculate the RDM of the system using the approach introduced in Ref.~\cite{Cirac2011, Schuch2013}. In order to efficiently contract the TN representation of the ruby RVB state, we use the simplex construction of the RVB state introduced in Sec.~\ref{sec:PEPS-RVB}-C. The main advantage of this construction is that the simplex tensors form a kagome TN which can be efficiently contracted around a cylinder using the BMPS technique introduced in Appendix~\ref{app:BMPS}.   

We calculated the EE of the RVB wave function in the even sector with winding number ($0,0$) on an infinitely-long cylinders of perimeter $N_v$ up to $N_v=8$ which was the maximum affordable cylinder width according to our available resources. Let us further note that the calculations, in general, can be done in any desired sector and will lead to the same result in the thermodynamic limit. However, from the technical point of view, implementation of the even sector is more convenient and less prone to error. However, other topological sectors can also be fixed by a specific choice of the boundary condition in the BMPS contractions. We refer the interested reader to Ref.~\cite{Poilblanc2012,Schuch2013,Iqbal2014} for a detailed discussion on the boundary conditions and on fixing the topological sectors.

Fig.~\ref{Fig:Svn} demonstrates the scaling of the EE as a function of $N_v$. One can clearly see that the EE of the RVB wave function obeys an area law with the width of the cylinder, $N_v$, which is best detected by the linear-fit (dashed line) in the figure and is in full agreement with the area-law term of $S=\alpha N_v-\gamma$. By setting $N_v=0$, we obtain the non-zero topological EE $\gamma=-\rm{ln}\ 2$ which is a clear signature of a $\Zd$ topological spin liquid with $\mc D=2$. Our findings reveal that the RVB wave function on the ruby lattice belongs to the same family of abelian topological phases as that of the ground state of the toric code \cite{Kitaev2003}.

 \section{Parent Hamiltonian} 
\label{sec:Par-Ham} 

In this section, we elaborate on possible parent Hamiltonians that the RVB wave function might serve as their ground state. The RVB states are potential candidates to be the ground state of the $\rm{SU(2)}$ quantum antiferromagnetic models \cite{Lacroix2013}. The $\Zd$ RVB state on the kagome lattice is a popular example which has already been suggested \cite{Iqbal2011} as a possible ground state for the spin-$\frac{1}{2}$ antiferromagnetic Heisenberg (AFH) Hamiltonian, i.e,
\be
\label{eq:H_AFH}
H = J \sum_{\langle i j \rangle} \mathbf{S}_i \cdot \mathbf{S}_j, 
\ee 
where the sum runs on all nearest-neighbors of the underlying lattice. We, therefore, explored the possibilities of the ruby RVB state being the ground state of the AFH model. To this end, we used the Hamiltonian~\eqref{eq:H_AFH} as a reference model and calculated the energy of the ruby RVB state~\eqref{eq:RVB1}. Using the CTMRG algorithm with boundary bond dimension $\chi=64$ to contract the infinite ruby lattice, we obtained $\varepsilon_0^{\rm RVB}=-0.378535$ for the variational energy of the RVB state in the thermodynamic limit. 

\begin{figure}  
  \centerline{\includegraphics[width=\columnwidth]{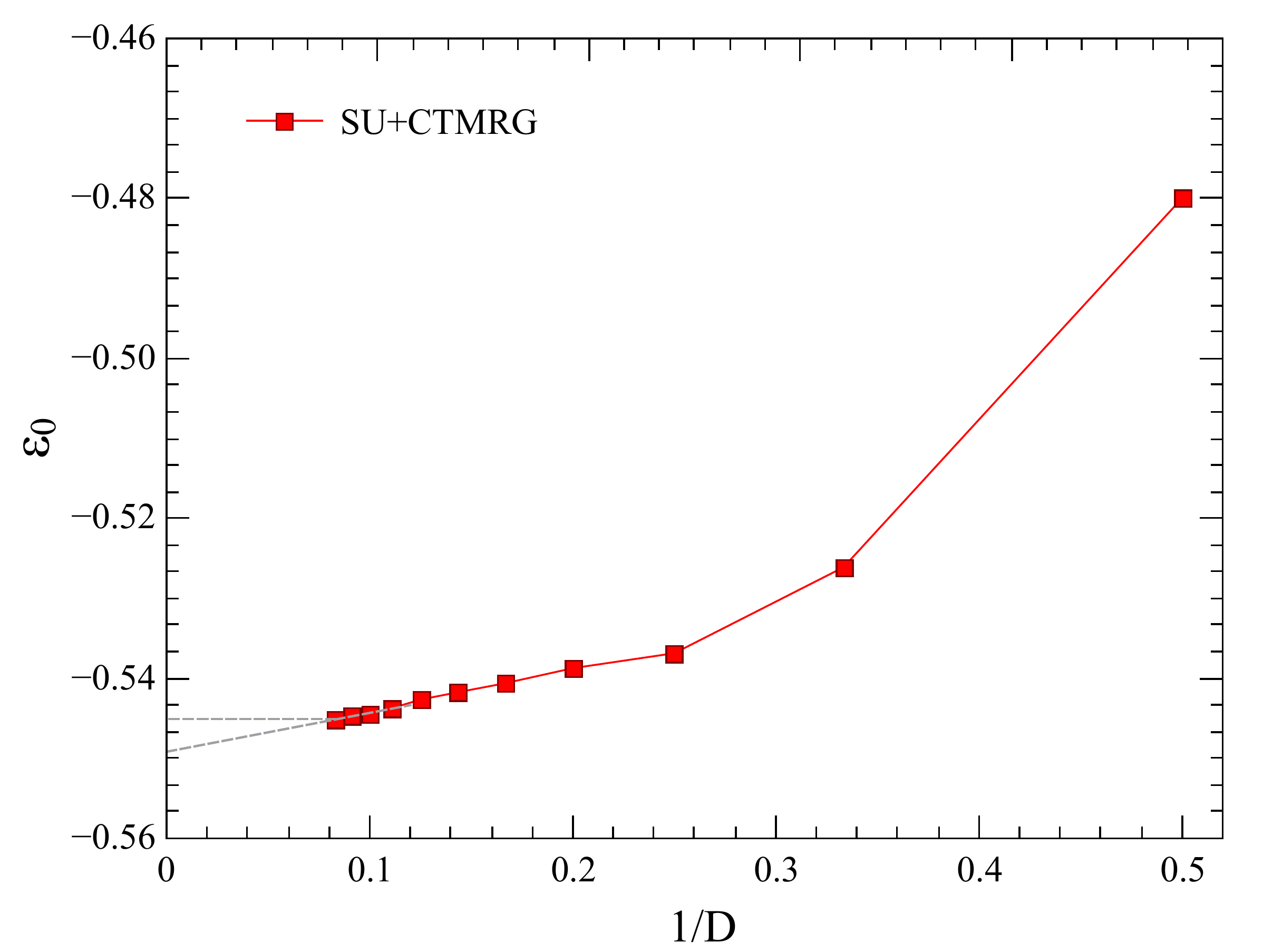}}
\caption{(Color online) Ground state energy per-site of the spin-$\frac{1}{2}$ antiferromagnetic Heisenberg model on the infinite ruby lattice with respect to inverse bond dimension, $1/D$, obtained with the iPEPS algorithm.}
\label{Fig:Eh}
\end{figure}

Next, by using the iPEPS algorithm previously developed for the simulation of the ruby lattice \cite{Jahromi2018}, we directly calculated the ground state wave function of the AFH Hamiltonian~\eqref{eq:H_AFH} up to $D=12$ in the thermodynamic limit. We used the simple-update of iPEPS based on imaginary-time evolution, accompanied by CTMRG for variational calculation of expectation values of local operators. Our best variational energy (obtained with iPEPS with $D=12$) for the AFH Hamiltonian on the ruby lattice is $\varepsilon_0^{\rm AFH}=-0.545089$ which is much below the RVB energy, $\varepsilon_0^{\rm RVB}$, suggesting that the ruby RVB is unlikely to be the ground state of the spin-$\frac{1}{2}$ antiferromagnetic Heisenberg model on the ruby lattice. 

\begin{figure}  
  \centerline{\includegraphics[width=0.8\columnwidth]{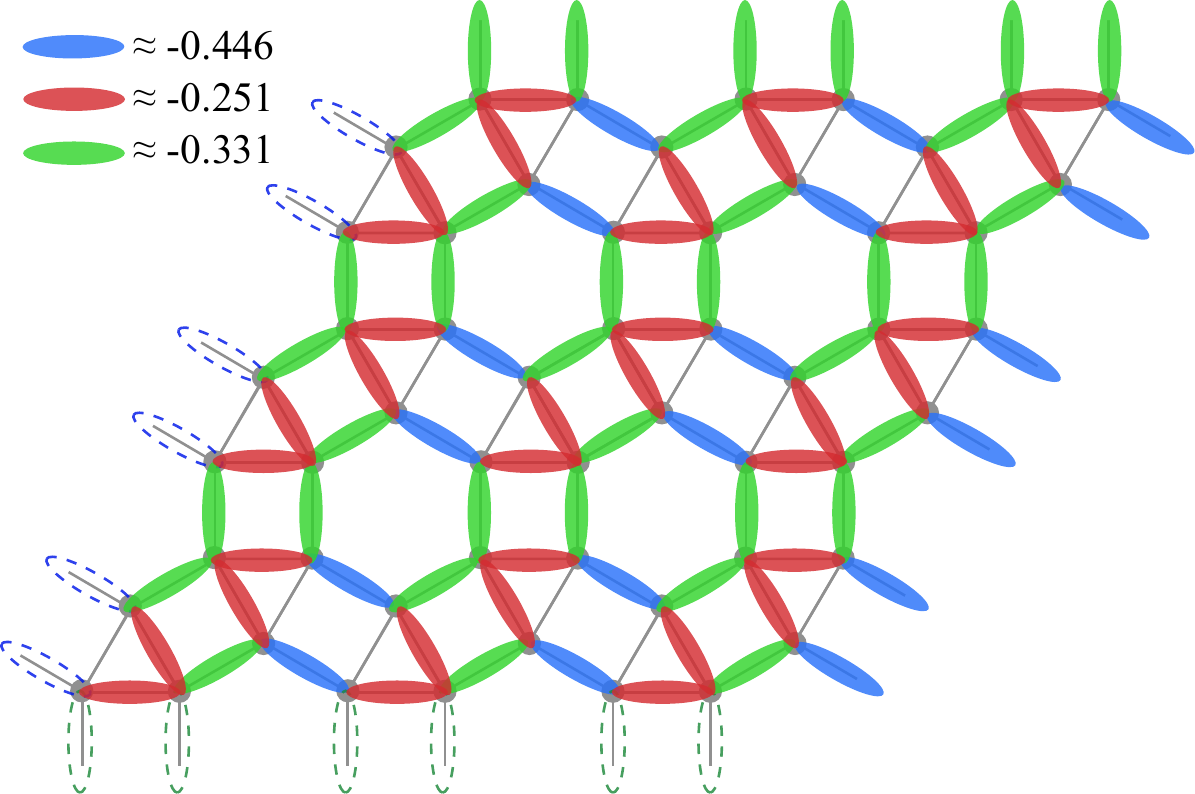}}
\caption{(Color online) The VBC ground state of the spin-$\frac{1}{2}$ AFH model on the ruby lattice composed of the decoration of three different dimers with different magnitude of NN spin-spin correlation, characterized in red, green and blue.}
\label{Fig:AFHOrder}
\end{figure}

Our results show that the ground state of the spin-$\frac{1}{2}$ AFH model on the ruby lattice is, in fact, a stabilized VBC phase characterized by the decoration of three different dimers of different NN spin-spin correlation (see Fig.~\ref{Fig:AFHOrder}). Let us further note that the VBC ground state of the AFH model on the ruby lattice is three-fold degenerate. Repeating the simulations with different initial states, we found the two other degenerate VBC states which are different from the others by a $C_3$ rotation of the dimer pattern around the big ruby plaquette. This is indeed in contrast with the RVB state which has a uniform distribution of spin-spin correlation all over the lattice with no broken symmetry.

The RVB state has also been predicted as the ground state of quantum-dimer models (QDM) on different lattices at the Rokhsar-Kivelson (RK) point \cite{Rokhsar1988,Moessner2001,Syljuasen2006,Moessner2002,Ralko2005}. Strong pieces of evidence have already been found to show that the ground state of the QDM on non-bipartite lattices is an RVB spin liquids \cite{Lacroix2013}. The ruby lattice is a non-bipartite structure. Besides, we have shown in previous sections that the ruby lattice already hosts a $\Zd$ RVB spin liquid. It is, therefore, reasonable to believe that a QDM Hamiltonian at its RK point can be the parent Hamiltonian for the ruby RVB state.  

In the following, we construct a QDM Hamiltonian for the ruby lattice. The building block of the ruby lattice is the diamond-plaquette, $\nores$, which paves the lattice and is also a ruby unit-cell (see also Fig.~\ref{Fig:ruby}-(b)). Due to the hard-core dimer covering constraint of the Hilbert space of the RVB state, such a ruby plaquette is commensurate only with two possible sets of local resonances given by \{$\resloopa,\resloopb$\} and \{$\resflipa,\resflipb$\}. One can, therefore, construct the following QDM Hamiltonian for the ruby lattice:
\bea
\label{eq:H-QDM}
H_{\rm QDM}&=&t\sum_{\nores}\bigg(\big|\resloopa\big\rangle\big\langle\resloopb\big|+\big|\resflipa\big\rangle \big\langle\resflipb\big|\bigg)+{\rm h.c.} \nonumber\\
	&+&v\sum_{\nores}\bigg(\left|\resloopa\right\rangle \big\langle\resloopa\big|+\big|\resloopb\big\rangle \big\langle\resloopb\big| \nonumber\\
	&+&\big|\resflipa\big\rangle \big\langle\resflipa\big|+\big|\resflipb\big\rangle \big\langle\resflipb\big|\bigg),
\eea
which acts on the diamond plaquettes and contains a kinetic and a potential term. The kinetic term is given by the first sum in the above Hamiltonian which is an off-diagonal term and creates resonances on the ruby lattice. The potential term, on the other hand, is given by the second sum which is diagonal and simply counts the number of plaquettes able to resonate, i.e., the flippable plaquettes. The $t$ and $v$ are the couplings tuning the strength of the kinetic and potential terms, respectively. The Hamiltonian~\eqref{eq:H-QDM} might serve as the parent Hamiltonian of the RVB state with an equal-weight superposition of all dimer-covering on the ruby lattice.

Let us further note that since the implementation of plaquette updates in the iPEPS machinery is a cumbersome and non-trivial task, we were not able to numerically test in what ranges of couplings, $t/v$, the system might lead to an RVB ground state. However, from the analogy with previous studies on other non-bipartite lattices such as the kagome \cite{Poilblanc2011,Misguich2002} and triangular lattices \cite{Moessner2002,Ralko2005}, the $t/v=1$ (RK point) is most likely to host the RVB wave function as its potential ground state. Last but not least, let us mention that there are more involved strategies based on the concept of injectivity \cite{Schuch2013,Schuch2010} in the PEPS language to construct generic parent Hamiltonians for families of tensor network states \cite{Schuch2012}.

 \section{Conclusions and discussion} 
\label{sec:conclude}

\begin{figure*}
\centerline{\includegraphics[width=10cm]{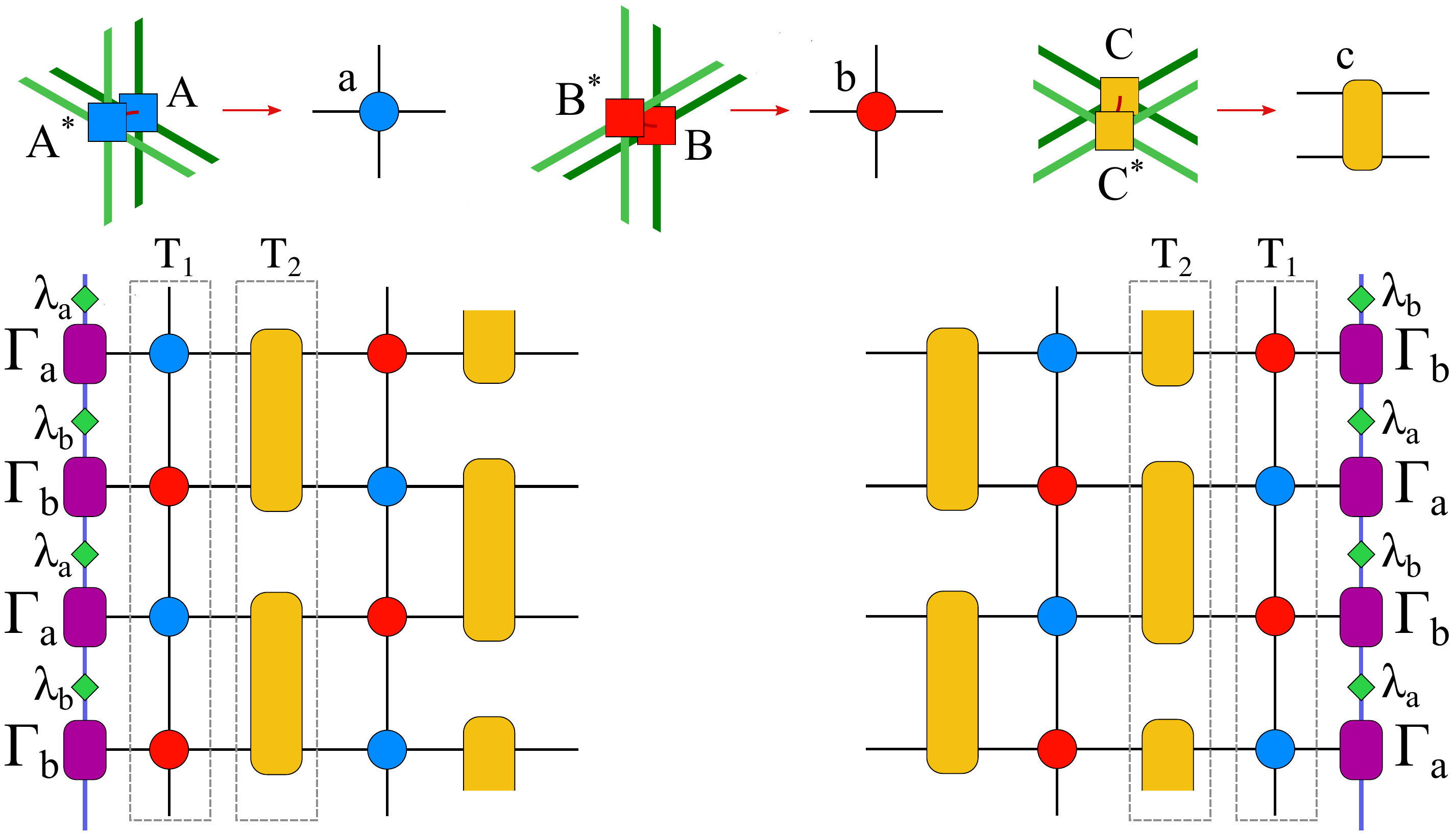}}
\caption{(Color online) (top) The kagome double tensors $a,b$ and $c$ which are obtained by contracting the local tensors and their conjugate along the physical legs and then grouping the corresponding virtual legs to form new rank-four tensors with virtual dimension $D^2$. (bottom) Left and right boundary MPS tensors, $\Gamma$, and the corresponding singular matrices $\lambda$, as well as the $T_1$, $T_2$ transfer operators which are absorbed to the left and right boundaries to obtain their left and right dominant-eigenvectors.}
\label{Fig:BMPS}
\end{figure*}

In this paper, we constructed a resonating valence-bond wave function composed of an equal-weight superposition of all possible nearest-neighbor hard-core dimer coverings on the ruby lattice. Our construction relied on the projected entangled-pair state formalism with bond dimension $D=3$. By introducing non-local moves to the dimer patterns, we found four distinct topological sectors on the torus which are distinguished by winding numbers defined according to the parity of the number of dimers crossed by non-contractible reference lines for every toroidal-direction. Next, we confirmed the topological nature of the ruby RVB wave function by calculating the entanglement entropy of a bipartition of the RVB state on the cylinder. We found the non-zero topological contribution to the entanglement entropy, $\gamma=-\rm{ln}\ 2$, which is consistent with that of a $\Zd$ topological spin liquid. 

Moreover, we examined the possibilities of the spin-$\frac{1}{2}$ antiferromagnetic Heisenberg model being the parent Hamiltonian for the RVB state on the ruby lattice. Our iPEPS simulation at large bond dimension revealed a VBC ground state with much lower energy for the AFH model compared to that of the RVB state on the ruby lattice, hence ruling out the AFH Hamiltonian as a candidate for parent Hamiltonian of the RVB state on the ruby lattice. Following our search for a relevant candidate parent Hamiltonian, we constructed a quantum-dimer model on the ruby lattice. We argued that the QDM on the ruby lattice might be a potential parent Hamiltonian for the RVB wave function at the Rokhsar-Kivelson point.  

Our RVB construction on the ruby lattice is of particular importance since it opens the door for future quest for topological spin liquids on SU(2) models on more complex structures. Let us remind that recently a U(1) QSL has been detected both experimentally and numerically on vanadium oxyfluoride compounds \cite{Clark2013,Aidoudi2011,Orain2017,Iqbal2018,Repellin2017}. Our topological RVB QSL construction can motivate further experimental studies on bismuth compounds such as $\rm{Bi_{14}Rh_3I_9}$ for the realization of new instances of QSLs. 

Most Importantly, the spin-orbit coupled Mott insulators with anisotropic interactions on the ruby lattice, such as the Kitaev model, already host topological phases as their low-energy effective theory. In particular, the Kitaev model on the ruby lattice at large $J_z$ coupling with dominant triangles represents a topological color code on the effective Honeycomb lattice which is $\Zd\times\Zd$ topologically ordered. In fact, the TCC is two copies of the famous toric code and belongs to the same Abelian topological phases. It has already been shown that the RVB state on the kagome lattice is adiabatically connected to the ground state of the toric code. We conjecture that our RVB wave function should also be connected to the ground state of the toric code on the ruby lattice. We leave its numerical proof as an open problem for future studies.    

Finally, in this paper, we developed and envisaged the machinery of the boundary MPS technique for the kagome lattice which can be served as an efficient algorithm for further investigation of quantum many-body wave functions, directly on the kagome lattice, and which can be used for calculation of physical quantities such as ground state fidelity and entanglement spectrum on kagome geometry.

\section{Acknowledgements}
The authors acknowledge fruitful discussions with Kai Philip Schmidt. The CPU time from ATLAS HPC cluster at DIPC is also acknowledged.   

\appendix

\section{Boundary MPS for the kagome lattice}
\label{app:BMPS}

In this appendix, we provide the details of the boundary MPS technique for the kagome lattice. This can be used for TN simulations such as the contraction of the infinite 2D kagome tensor network on a cylinder.

Before we continue with details of the BMPS, let us remind that the kagome lattice has a translationally invariant three-site unit-cell. Although there is a freedom in the choice of the three-site unit-cell which may vary according to the specific technique, here we consider a triangular unit-cell such as the one shown in Fig.~\ref{Fig:block}-(c,d) with three different tensors.

Let us note that we use the double TN contraction scheme in our BMPS machinery, i.e, we first contract the local tensors for the ket and bra (it's conjugate) along their physical legs and then group the virtual legs of the ket and the bra tensors to form local double tensors of rank-four with virtual dimension $D^2$. Fig.~\ref{Fig:BMPS} (upper half) illustrates the three double tensors $a,b$ and $c$ in a kagome unit-cell. Contraction of the kagome TN from the left and the right can then be represented as the left and right dominant-eigenvector of the transfer operators $T_1$ and $T_2$. 

In order to approximate the left and right dominant-eigenvectors, we start from two sets of boundary tensors \{$\Gamma_a,\Gamma_b$\} and two sets of diagonal matrices for the boundary singular values \{$\lambda_a,\lambda_b$\} for both left and right boundaries and start the BMPS procedure by iteratively applying with $T_1$ and $T_2$ transfer operators on boundary MPS tensors. One should note that by contracting the $T_1$ and $T_2$ transfer operators, which are nothing but the double tensors $a,b$ and $c$, the dimensions of the virtual bonds of the boundary MPSs $\Gamma$ grows exponentially and therefore, we have to truncate the virtual bonds back to some desired fixed value $\chi$, which is called boundary bond dimension. The truncation is done locally by taking the singular-value decomposition (SVD) and keeping the $\chi$ largest singular values and discarding the rest \cite{Vidal2007,Vidal2007a}.

Since the transfer operators $T_1$ and $T_2$ are not unitary, the boundary MPS tensors will not be in the canonical form and the singular values will misbehave after several iterations leading to failure in SVD convergence. To resolve this issue, we need to manually bring the boundary MPS tensors into the canonical form. Several methods have already been suggested for the canonicalization of arbitrary MPSs which have their own benefits and downsides \cite{Vidal2007a,Orus2008}. In what follows, we use the iterative technique introduced in Ref.~\cite{Kalis2012}. The basic idea is actually very simple. In this approach, after absorbing each gate (transfer operators) to the boundary we do not immediately truncate to $\chi$ but rather, we sweep over the whole MPS tensors and iteratively join every two neighboring MPS tensors ($\lambda_a,\lambda_b$) along their connected legs and split it by SVD. This process is like acting with the gate identity on the boundary MPS tensors. We repeat this process several times ($10-20$ iteration) and at the final iteration, we truncate the singular values back to $\chi$.   

After obtaining the converged left and right boundary MPS tensors, we can use them for different purposes such as calculating the boundary density operators and entanglement spectrum as well as the entanglement entropy, as discussed in Ref.~\cite{Cirac2011, Schuch2013}. 

Finally, let us stress that the correct choice of the initial boundary MPS tensors is of crucial importance. In particular, the topological sector of a topological wave function such as the RVB state is fixed at the boundary \cite{Poilblanc2012,Poilblanc2013} and a choice of irrelevant random boundary MPS tensor might lead to loss of topological information or underestimation of topological entropy. For example, we observed in our simulation that the correct topological contribution can be obtained correctly only if we put the RVB MPS tensors on the boundaries.

\bibliography{references}{}
\bibliographystyle{apsrev4-1}

\end{document}